\documentclass[aps,prl,twocolumn,groupedaddress,showpacs,showkeys]{revtex4-1}
\usepackage{amsmath,amssymb,amsthm,graphicx,fancyhdr,epsf,latexsym,epstopdf,psfrag}
\usepackage[usenames,dvipsnames]{color}

\newcommand{\ab}{\textit{ab initio }}
\newcommand{\abb}{\textit{Ab initio }}

\begin{document}

\preprint{AIP/123-QED}

\title{Insights on finite size effects in \abb study of CO adsorption and dissociation on Fe 110 surface }
%\thanks{Footnote to title of article.}

\author{Aurab Chakrabarty$^1$}
\email{aurab.chakrabarty@qatar.tamu.edu}

\author{Othmane Bouhali$^1$}
\author{Normand Mousseau$^2$}
\author{Charlotte S. Becquart$^3$}
\author{Fadwa El Mellouhi$^{4}$}

\affiliation{$^1$Texas A\&M University at Qatar, P.O. Box 23874, Doha, Qatar}
\affiliation{$^2$D\'epartement de physique and RQMP, Universit\'{e} de Montr\'{e}al, Case postale 6128, succursale centre-ville, Montr\'{e}al (QC) Canada H3C 3J7}
\affiliation{ $^3$UMET, UMR CNRS 8207, ENSCL, Universit\'{e} Lille I, 59655 Villeneuve d'Ascq c\'{e}dex, France}
\affiliation{$^4$Qatar Environment and Energy Research Institute, P.O. Box 5825 Doha, Qatar}

\date{\today}% It is always \today, today,%  but any date may be explicitly specified

\begin{abstract}
Adsorption and dissociation of hydrocarbons on metallic surfaces represent crucial steps on the path to carburization, eventually leading to dusting corrosion. While adsorption of CO molecules on Fe surface is a barrier-less exothermic process, this is not the case for the dissociation of CO into C and O adatoms and the diffusion of C beneath the surface, that are found to be associated with large energy barriers.
In practice, these barriers can be affected by numerous factors that combine to favour the CO-Fe reaction  such as the abundance of CO and other hydrocarbons as well as the presence of structural defects. From a numerical point of view, studying these factors is challenging and a step-by-step approach is necessary to assess, in particular, the influence of the finite box size on the reaction parameters for adsorption and dissociation of CO on metal surfaces. Here, we use density functional theory (DFT) total energy calculations with the climbing-image nudged elastic band (CI-NEB)  method to estimate the adsorption energies and dissociation barriers for different CO coverages with surface supercells of different sizes. We further compute the effect of periodic boundary condition for DFT calculations and find that the contribution from van der Waals interaction in the computation of adsorption parameters is important as they contribute to  correcting the finite-size error in small systems. The dissociation process involves carbon insertion into the Fe surface causing a lattice deformation that requires a larger surface system for unrestricted relaxation. We show that, in the larger surface systems associated with dilute CO-coverages, C-insertion is energetically more favourable, leading to a significant decrease in the dissociation barrier. This observation suggests that a large surface system with dilute coverage is necessary for all similar metal-hydrocarbon reactions in order to study their fundamental electronic mechanisms, as an isolated phenomenon, free from finite-size effects.
\end{abstract}

\pacs{68.43.Bc, 66.30.Jt, 68.47.De}% PACS, the Physics and Astronomy Classification Scheme.
\keywords{Periodic boundary condition, Carburization, Metal dusting, Metal-hydrocarbon reaction, bcc Fe, CO adsorption, Dissociation }%Use showkeys class option if keyword
                  
\maketitle

\section{\label{sec:level1}Introduction}

Metal dusting is a severe form of corrosion that plagues the petrochemical industry when iron is in contact with hydrocarbons at high temperature. The primary step of the dusting corrosion process is the carburization of metal surface, that is commonly described by the Fischer-Tropsch (F-T) synthesis. Hydrocarbons react with transition metals at elevated temperature, in a so-called carburizing atmosphere where the carbon activity constant, a$_c$, is much larger than 1. Between around 700 and 1100~K, CO and HCOH are adsorbed and dissociated on the Fe surface \cite{Zhang11,Pippel98}. C diffuses into the metal surface while H reacts with O to produce steam. As the carburization process goes on,  C atoms deposits accumulate in the metal to form a metastable cementite (Fe$_3$C) phase. At this point, Fe atoms diffuse into cementite and erupt as Fe metal dust \cite{Young11,Pippel98}. As carburization is the initiating step in the metal dusting corrosion (MDC) and is also important for cutting the emission of greenhouse gases CO and CO$_2$, it has been studied widely both experimentally and numerically, with the use of various computational approaches \cite{Zhang11,Zhang12,Jiang04,Johnson10}. The reaction  between CO and Fe surface can be broken into two consecutive processes, adsorption and dissociation. Adsorption describes the process of a CO molecule getting attached to the Fe surface. During dissociation, this molecule decomposes into C and O adatoms directly at the surface, with the two atoms moving into their most stable adatom site. Adsorption of CO is a barrier-less exothermic process. We show here, confirming earlier first-principles study \cite{Jiang04, Stibor02}, that adsorption releases an energy near 2 eV/molecule and depends heavily on the surface indexes and the symmetry of the adsorption site. Vibrating-capacitor experiments suggest an adsorption energy of 1.24 eV \cite{Wedler83}.  Dissociation, on the other hand, is a costlier step that requires breaking a C-O triple bond, with a 11.2 eV/molecule average dissociation energy in vacuum \cite{Sanderson83}. The metal surface catalytic effect must play therefore a crucial role for breaking this bond \cite{Young11}. Earlier density functional studies show that on Fe-110 surface, the CO dissociation barrier is reduced to 1.5 eV \cite{Jiang04}. Recent studies on 100 \cite{Wang14} and 111 \cite{Booyens14} surfaces also show similar adsorption and dissociation energies. However, a number of experiments suggest  CO is adsorbed on the 110 surface at low temperature with dissociation taking place at higher temperatures with an onset observed at at least 380~K  \cite{Wedler82, Wedler83}.  Therefore with a barrier of 1.5 eV, CO dissociation on a clean Fe surface has only a negligible probability from a thermal activation perspective. In practice, more complex reaction pathways would be possible as a number of other molecules and sites that could act as activator, such as H and other reactive hydrocarbons and surface defects, are generally present and could assist the dissociation process \cite{Elahifard12, Tian14}. Recent works on the 100 surface show that high-coverage presence of H increases the rate of carburization by reducing the dissociation barrier \cite{Roncancio15, Elahifard12}. On the other hand Cu impurities on the surface increase the dissociation barrier and prevent C atoms to be adsorbed into the surface \cite{Tian14}.

\begin{figure}
\centering
\includegraphics[width=4cm]{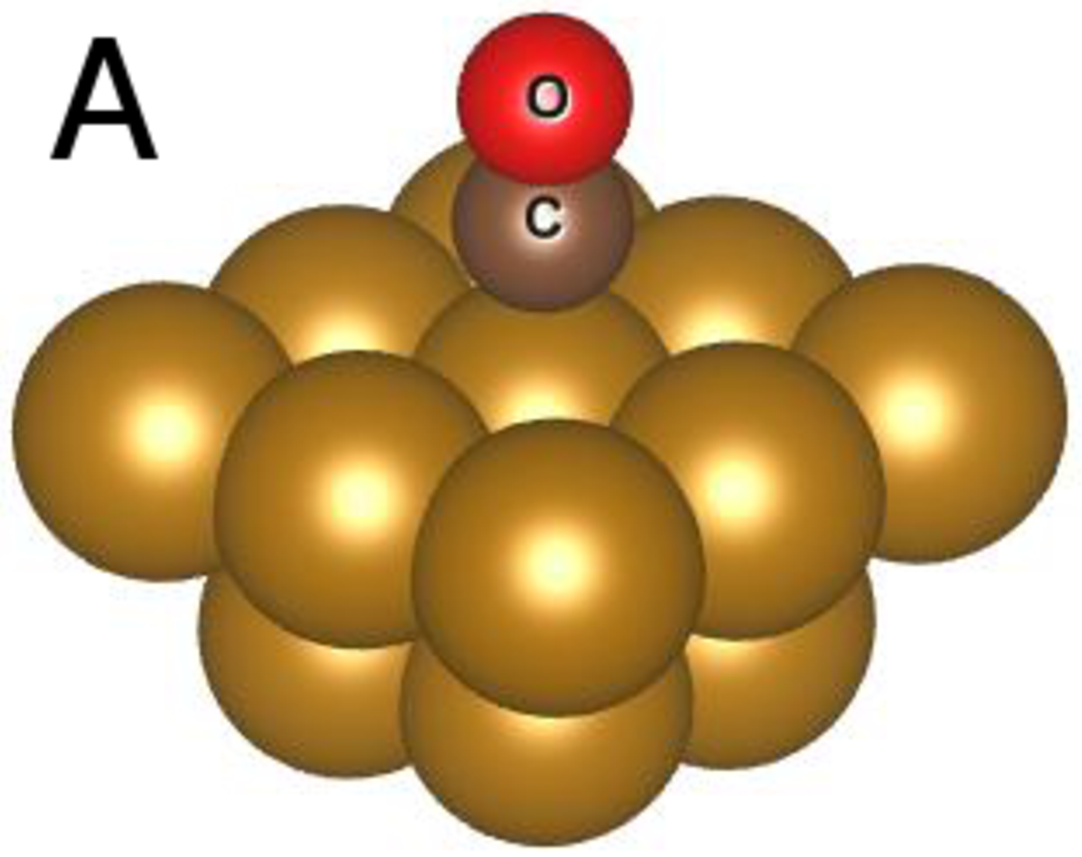}
\includegraphics[width=4cm]{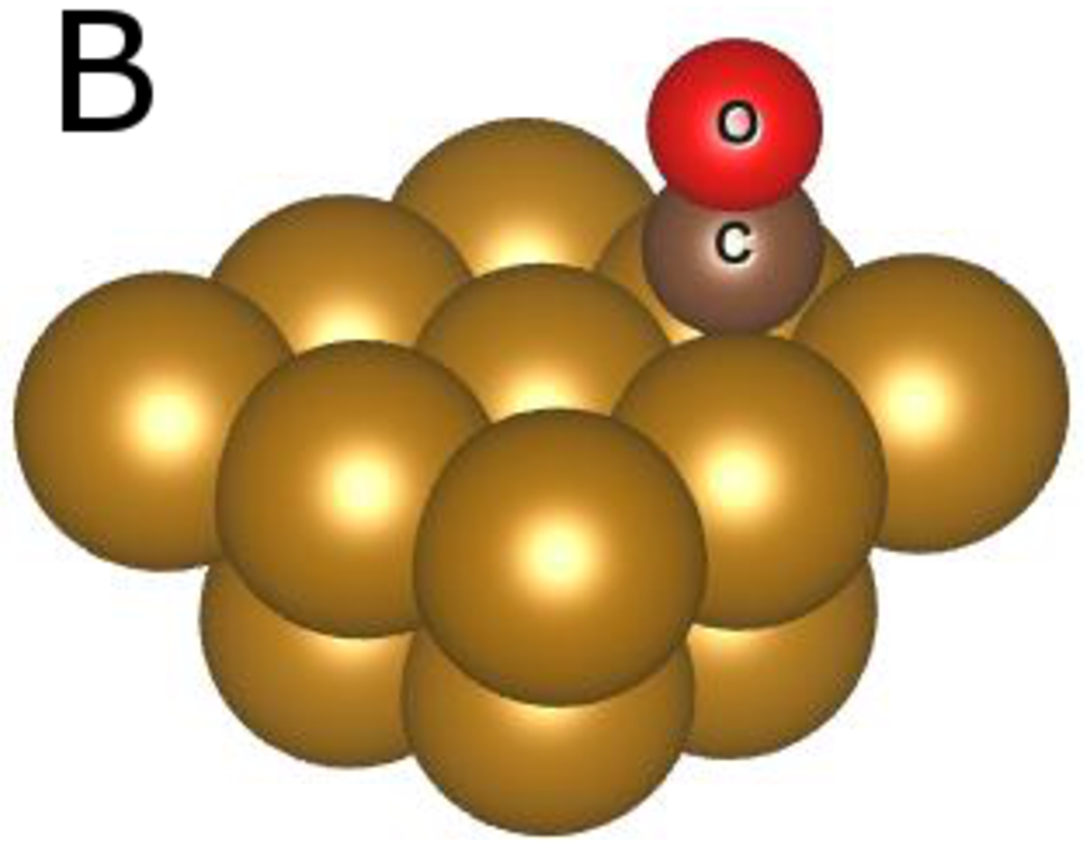}
\includegraphics[width=4cm]{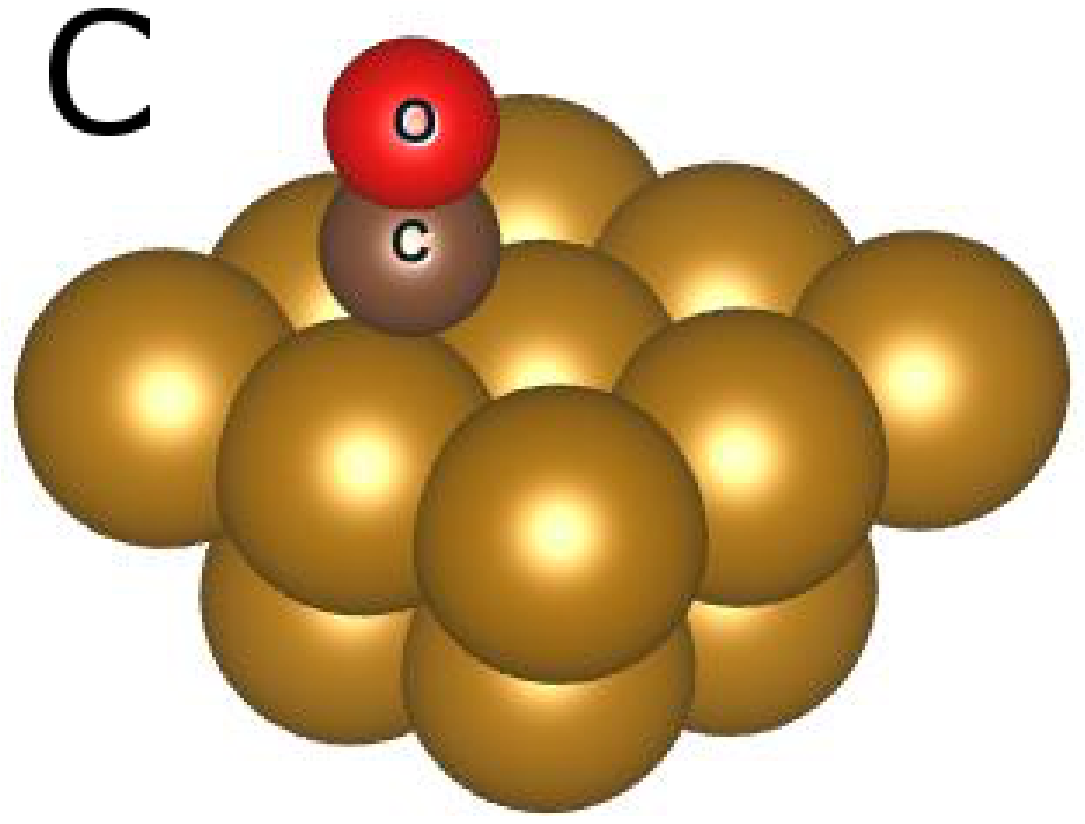}
\caption{Different adsorption sites for CO on Fe-110 surface. A: on-top, B: hollow and C: bridge site.}
\label{sites}
\end{figure}

Experimental studies using low energy electron diffraction (LEED) and high-resolution electron energy loss spectroscopy (HREELS) suggest an upright orientation of CO molecule when adsorbed to one of the possible high-symmetry sites on-top, hollow and bridge, shown in Fig. \ref{sites}. However, at very high coverage, a tilted orientation is preferred \cite{Wang14}. Energetically, at a coverage as high as 0.5 monolayers (ML), a hollow site is preferred for adsorption while at a smaller coverage of 0.25 ML an on-top site is  more stable \cite{Jiang04, Stibor02}. Due to the small differences between the adsorption energies, computational precision is important in calculation. However, semiempirical tight-binding model calculations \cite{Mehandru88} and earlier DFT calculations \cite{Hammer99} show contradictory  results regarding the most stable adsorption site. A literature survey of more recent works, that use standard GGA PW91 \cite{Perdew92} and PBE \cite{Perdew96} functionals and meta-GGA \cite{Janesko15}, PKZB \cite{Perdew99}, or RPBE \cite{Hammer99} approximations, reveal that the exact adsorption energy on particular sites depends on the choice of the exchange-correlation functional \cite{Kurth99, Jiang04, Booyens14}. For example, studies of CO adsorption and dissociation on 110 \cite{Jiang04} and 111 \cite{Booyens14} Fe surfaces using PBE, PZKB and RPBE methods show that the meta-GGA functional does not provide any significant qualitative advantage over standard PBE-GGA. Moreover, while perturbation techniques such as MP2 are excellent in describing organic molecular systems, they show diverging behavior for metallic solids \cite{Shepherd12}. Hybrid Hartree-Fock functionals are also generally rejected in the study of metallic systems because of inaccurate description of these systems and high computational cost in a plane-wave platform. On the other hand PBE-GGA tends to yield a consistent relative energy between adsorption energies at different sites, in agreement with a number of experimental parameters such as vibrational frequencies and work function changes for CO adsorption \cite{Jiang04, Booyens14}. 

Jiang \textit{et al} \cite{Jiang04, Jiang05} carried out extensive first-principles DFT studies of adsorption and dissociation of CO on Fe surfaces in 2004-2005, using the PBE-GGA exchange-correlation functional with the nudged elastic band (NEB) \cite{Henkelman00} method to converge diffusion pathways, an approach followed by a number of groups \cite{Zhang11, Zhang12, Booyens14}. After almost a decade, this topic has been revisited by many researchers in recent years \cite{Elahifard12, Booyens14, Wang14, Tian14, Roncancio15, Wang14a}. There are several reasons for this renewed interest, of which, notable from the perspective of computational science, is the advancement in computational power that is necessary for surface calculations with the computationally demanding nudged elastic band method. Therefore not only the simulations of large surfaces with dilute CO-coverages are feasible, but also the influence of impurities and defects on these processes can be observed. None of these later studies, however, used a dilute coverage with a focus on the electronic mechanisms of the adsorption and dissociation of CO as isolated processes. In earlier first-principles and experimental studies, 0.25 ML was the lowest coverage studied \cite{Wang14, Stibor02, Booyens14,Jiang04, Wedler82, Wang14a}. Only higher coverages were studied due to the fact that onset of carburization was postulated at a dense hydrocarbon environment that represents a high carbon activity ($a_c > 1$). The system-size limitations for \ab calculations also prevented  more costly calculations with large supercells that is required for dilute coverages.  A recent study on the specific problem of coverage-dependence of CO reaction with various Fe surfaces show that the dissociation barrier is reduced for lower coverages up to 0.25 ML \cite{Wang14a}. A study on 100 surface show that near saturated coverage (0.7-1 ML) desorption of CO becomes energetically more favorable than dissociation \cite{Wang14}.

With more advanced computational power and efficient parallel processing algorithms, some of these unanswered questions regarding the reaction of CO with the Fe-110 surface can now be addressed. In order to understand the atomistic mechanism of CO adsorption and dissociation, it is necessary to first characterize it as an isolated process, excluding interaction with other adsorbing / dissociating CO molecules. This is possible by choosing a low coverage, \textit{i.e.} a large surface area. A large surface area makes computation expensive but is affordable with modern computers and is essential if the interactions associated with periodic boundary condition are to be avoided, allowing a better description of the reaction mechanism between CO and Fe surface as an isolated process. 

In this work we study the adsorption and dissociation of CO as functions of increasing box sizes corresponding to coverages from 0.25 ML to 0.0625 ML using PBE-GGA exchange correlation functional. With the 0.25 ML coverage model, we first establish the reliability of our computational setup by reproducing adsorption and dissociation results obtained in previous publications. With the lower coverage models, we observe effect of periodic boundary condition on the CO-Fe carburization reaction, showing that the molecular interaction between the CO molecule and its periodic images is not complete without considering van der Waals interactions. We show, in particular, that including van der Waals interactions makes a significant difference in the adsorption energies of CO at high coverages when the distance between CO molecules are small. For the dissociation process, the impact of periodic boundary condition is even greater. We show that when the coverage is decreased from 0.25 ML to 0.0625 ML, and the dissociation barrier is significantly reduced from 1.64 eV to 0.78 eV. This implies that the CO-Fe dissociation reaction becomes more favorable at lower coverages.
 In particular, we assess in details the various factors that govern the relation between the dissociation barrier and CO-coverage, which leads us to conclude that upon carburization, the high-coverage systems suffer from uncompensated elastic deformation.
In the first-principles calculations, the high-coverage systems are modeled with small surface supercells that, due to the periodic image interaction under boundary conditions, do not allow an unobstructed relaxation upon CO dissociation, therefore elevating the transition state energy.
The novelty of this work is to use a large 4$\times$4 supercell representing a very dilute coverage of 0.0625 ML, such that the fundamental electronic mechanism of adsorption and dissociation can be observed as an isolated phenomenon, free from periodic image interaction.
 The observation made here with the dilute coverage has a serious impact on the wide field of studies on fundamental electronic mechanism of the metal-hydrocarbon reactions as well as on the influence of other adsorbates on these reactions. Our study suggests the necessity of using a large surface / dilute coverage system for the sake of better understanding of the fundamental processes, as well as influence of other adsorbates or defects, as isolated phenomena, before conducting these studies using a high coverage / small supercell systems.

\section {\label{sec:level1} Computational Methods}

DFT calculations are performed using spin-polarized PBE-GGA exchange-correlation functional with the plane-wave basis pseudopotential method implemented in the Vienna \ab simulation package (VASP) \cite{VASP}. Total energies are calculated with a kinetic energy cutoff fixed at 400 eV for all calculations. Grimme's DFT-D3 \cite{Grimme10} method was used to incorporate van der Waals interactions. Convergence criteria of 10$^{-4}$eV/atom and 10$^{-3}$\AA~were selected for total energy and structure relaxation, respectively. A ferromagnetic (FM) bcc structure with space group $Im\bar{3}m$ was used for bulk Fe. The equilibrium bulk lattice constant was found to be 2.83148\AA.

We first look at the reconstruction of the three
lowest index surfaces, 100, 110 and 111, to test the accuracy of our results as well as to
find out a suitable surface for the CO-Fe reaction. Supercells are created using a
$n\times n\times 10$ simulation box, where $n$ is the multiplicity factor of the unit
cell in the dimensions parallel to the surface. Therefore, in terms of the lattice
vectors associated with the surface unit cells $a$, $b$ and $c$, with $c$ being perpendicular
to the surface, the box size is $na\times nb\times 10c$. The box is filled up in such a
manner that there is 9-10 {\AA} of vacuum. Previous works with PAW-PBE pseudopotential
method confirmed that a vacuum of 10 {\AA} is adequate as the interaction between slabs
due to the periodic boundary condition is negligible \cite{Blonski07,Spencer02}. The
100, 110 and 111 surface supercells have 13, 13 and 17 layers respectively and therefore
the corresponding vacuum lengths are 11.4, {\AA}, 16 {\AA}, and 9.8 \AA, respectively.
 The top five layers of each surface are free to relax in all 3 dimensions.
 Three further layers are free to relax only in the direction perpendicular to the surface ($c$).
 The rest of the supercell atoms are kept fixed in their ideal position to simulate an infinite bulk.

 For all the adsorption and dissociation calculations we used primarily three different supercells for the 110 surface,  i.e. 2$\times$2,   3$\times$3  and  4$\times$4 multiples of the 110 surface unit cell, all of which are 8 layers thick with 16 \AA{} of vacuum space. A dense K-point mesh is automatically generated using Monkhorst-Pack grid.  For the adsorption calculations 12$\times$12$\times$1, 9$\times$9$\times$1, and   7$\times$7$\times$1 grids were used for 2$\times$2, 3$\times$3 and 4$\times$4 surfaces respectively. For the calculations of the minimum energy path for dissociation a coarser set of 9$\times$9$\times$1, 7$\times$7$\times$1, and  5$\times$5$\times$1 grids are used. Two additional surface supercells 2$\times$3, and 3$\times$4 are also considered in the study of the finite size effects. 12$\times$9$\times$1 and 9$\times$7$\times$1 k-point grids are used for the  2$\times$3, and 3$\times$4 surface supercells, respectively.

The surface energy $E_{\mathrm{surf}}$ is calculated using,
\begin{eqnarray}
E_{surf} = \frac{E_{\mathrm{TOT}}(\mathrm{surface}) - NE_{\mathrm{TOT}}(\mathrm{bulk/atom})}{2A_{\mathrm{surf}}}
\label{Esurf}
\end{eqnarray}
where N is the number of atoms in the surface supercell and A$_{surf}$ is the surface area.

The adsorption energy is calculated using,
\begin{eqnarray}
E_{ads}^S =  E^{CO-surf}_{TOT} - E^{cln}_{TOT} (CO)
\label{EAds}
\end{eqnarray}
where $E^{CO-surf}_{TOT}$ and $E^{cln}_{TOT} (CO)$ are the total energies corresponding to a CO-adsorbed surface and a clean surface with a CO molecule placed in the vacuum away from the surface, respectively. Here, however, the self-energy of a standalone CO molecule is not calculated in an attempt to minimize computational error.

The climbing-image nudged elastic band (CI-NEB) method \cite{Henkelman00} implemented in VASP is used to trace the minimum energy paths (MEP) associated with the dissociation of CO. In order to find a transition state, the end-point (initial and final) configurations for the reaction are fully relaxed with high precision. NEB uses  an interpolated chain of intermediate configurations between the two end point configurations, connected by springs. The whole chain is then relaxed simultaneously with a fixed spring constant until the total average force minimizes under the tolerance limit of 0.01eV/\AA. In this work, 15 intermediate configurations or images are considered, starting from an on-top adsorbed to a fully dissociated configuration. The MEPs are obtained with 3-stage high-precision relaxation of the path. First, a basic MEP was obtained using conventional NEB. Next, climbing image method was applied on this primitive path with a fewer number of images (typically 5 per path) to fine tune the energy and position of the saddle point. Finally, we add intermediate points in the path obtained with climbing image to ensure that no small detail of the reaction path is missed and the complete path is relaxed again for a third time.

\subsection  {\label{sec:level2}Surface reconstruction: the choice of the system}

To characterize the three reconstructed surfaces, we consider their energy defined in Eq. \ref{Esurf} as well as the surface reconstruction parameter that describes the damping of the displacement of surface atoms  to convergence. This parameter $\delta z_n$, is the change in interlayer spacing of the n$^{th}$ layer from the surface, as a percentage of bulk interlayer spacing. Table \ref{Surf_recon} shows the complete set of surface reconstruction parameters calculated for the 100, 110 and 111 surfaces. A number of results from previous works are also cited for comparison.

%%%%%%%%%%%%%%%%%%%%%%%%%%%%  TABLE SURFACE RECONSTRUCTION
\begin{widetext}

\begin{center}
\begin{table}[h!t!p!]
\centering
\caption{\label{Surf_recon} Surface reconstruction for 100, 110 and 111 surfaces, the surface reconstruction parameter $\delta z_n$, as defined in the text, is compared with previous works.}
\vspace{2mm}
\setlength{\tabcolsep}{2mm}
\centering
\begin{tabular}{llccccc}
\hline
\hline
Surface	&  	Reference		& $\delta z_1$ &	$\delta z_2$	& $\delta z_3$ &   $\delta z_4$   &  $\delta z_5$  \\
\hline
                                                           
		&	This work 				&      -1.14	&   +2.08	&    +0.41	&     -0.07	&	-0.11	\\
100		&	PBE-GGA \cite{Blonski07} 		&      -3.6	&    +2.3	&     +0.4	&    -0.4	&	-0.5	\\
		&	PW91-GGA\cite{Spencer02} 	&	-1.89	&   +2.59	&     +0.21	&    -0.56	&	-0.14	\\
		&	Expt (LEEDS) \cite{Wang87}	& -5$\pm$2	& -5$\pm$2	&	-	&	-	&	-	\\
\hline
		&	This work				&     +0.04	&    +0.11	&	-0.22	&	-0.08	&       +0.03	\\
110		&	PBE-GGA \cite{Blonski07}		&	-0.1	&    +0.3	&	-0.5	&	-0.2	&       +0.04	\\
 	  	&	PW91-GGA\cite{Spencer02}	&	-0.13	&   +0.197	&      -0.006	&	-	&	-	\\
		&	Expt. (LEEDS) \cite{Wang87}	& +0.5$\pm$2	&   -	&    	-	&	- 	&	- 	\\
\hline
		&	This work				&     -12.65	&    -9.75	&      -2.91	&   -5.85	&	-0.63	\\
111		&	PBE-GGA \cite{Blonski07}		&     -17.7	&    -8.4	&      +11.0	&   -1.0	&	-0.5	\\
		&	PW91-GGA \cite{Spencer02}	&     -13.3	&    -3.6	&      +13.3	&   -1.2	&	+0.35	\\
		&	Expt. (LEEDS) \cite{Wang87}	& -16.9$\pm$3 & -9.8$\pm$3 &	+4.2$\pm$3.6 &	-2.2$\pm$3.6	&	-	\\ 
\hline
\end{tabular}

\end{table}
\end{center}
\end{widetext}

%%%%%%%%%%%%%%%%%%%%%%%%%%%%%%%%%%%%%%%%%%%%%%%%%

%%%%%%%%%%%%%%%%%%%%%%%%%%%%  TABLE SURFACE ENERGY
\begin{table}[ht]
\caption{\label{Surf_eng} Surface energies of 100, 110, and 111 surfaces in J/m$^2$, compared with previous works}
\vspace{3mm}
\setlength{\tabcolsep}{3mm}
\centering
\begin{tabular}{lccc}
\hline
\hline
Reference				& 	100	 &	110	&	111	 \\
\hline
This work (PBE-GGA)		&      2.50	&   2.44	&    2.80	\\
PBE-GGA \cite{Blonski07} 		&      2.47	&   2.37	&    2.58	\\
PW91-GGA\cite{Spencer02} 	&      2.29  	&   2.27	&    2.52	\\
MM \cite{Grochola02}		&      1.51	&   1.36	&    1.68	\\
MM \cite{Rodriguez93}		&      3.35	&   1.70	&    2.39	\\
Expt. (0K)\cite{Tyson77}		&      2.41	&   2.41	&    2.41	\\
Expt. (Melting) \cite{Tyson77}	&      2.12	&   2.12	&    2.12	\\
 
\hline
\end{tabular}

\end{table}

%%%%%%%%%%%%%%%%%%%%%%%%%%%%%%%%%%%%%%%%%%%%%%%%%

Table \ref{Surf_recon} shows that the displacement is relatively small except for the 111 surface, which has the lowest surface density of the three surfaces, and for which the first two layers are heavily contracted. This elastic relaxation drops below 1\% (equivalent to 0.02 \AA) at or before the depth of 5$^{th}$ layer for all the surfaces. As expected, the 110 surface, with its densely-packed structure, shows the least amount of reconstruction.

The surface energies, calculated using Eq.~\ref{Esurf}, are given in Table \ref{Surf_eng} and are compared with results from selected previous works. The values calculated in this work agree well with earlier first-principles works. The experimental result, which is estimated from a solid-liquid interfacial free energy technique \cite{Tyson77}, cannot distinguish between similar-index surfaces, but the average value is not far from the first-principles result when measured at $\sim$0K. 
As with the previous calculation results shown in  Table \ref{Surf_recon} and \ref{Surf_eng}, our simulation confirms that the 110 surface has the lowest reconstruction and surface energy. It is also the most densely packed surface given that the surface densities of 100, 110 and 111 surfaces are 1/\textbf{a}$^2$, $\surd2$/\textbf{a}$^2$ and 1/$\surd$3\textbf{a}$^2$, respectively, where \textbf{a} is the lattice parameter of bulk bcc Fe.

\section {\label{sec:level1}Results}

%%%%%%%%%%%%%%%%%%%%%%%%%%%%%%%%%%%%%

\subsection{Adsorption}

For the adsorption of CO molecule on the 110 surface, three possible symmetrical adsorption sites are considered: the on-top, hollow and bridge sites, all shown in Fig. \ref{sites}. Table \ref{Ads_eng} shows the adsorption energies for different sites at 0.25 ML , 0.1111 ML and 0.0625 ML coverages, corresponding to a single molecule on the 2$\times$2, 3$\times$3  and a 4$\times$4 surface systems, respectively. The adsorption energies for the 0.25 ML is in agreement with  Jiang \textit{et al}'s results \cite{Jiang04}. A negative sign indicates energy gain through adsorption (Eq. \ref{EAds}).  An on-top position is energetically the most favourable for adsorption at 0.25 ML of coverage and lower. The CO molecule sits upright with the C atom closer to the Fe surface (Fig. \ref{sites}) in all three symmetric sites. The C-O bond (1.17 {\AA} in vacuum) is stretched by 4.5 - 9.1\% with the largest stretch for the hollow site. The CO molecule relaxes to a position where the C atom is 0.9-1.1 {\AA} above the surface. One can see in Table \ref{Ads_eng} that the inclusion of van der Waals interactions lowers the adsorption energy and is drastic for the higher coverage, 0.25 ML. At 0.0625 ML mutual distances between CO molecules is larger, at 9.8 {\AA}, so that  the van der Waals interactions are relatively weak and the corrections due to this term in the adsorption energies are limited to 20-50 meV.
 Interestingly, the inclusion of van der Waals interactions brings adsorption energies for different coverages closer to each other. We note that careful selection of the dispersion correction method is crucial as there are reports of frequent overestimation of van der Waals interaction energies \cite{Ramalho13}. However, recent benchmarking density functional studies show that the molecular adsorption on metal is generally described correctly with van der Waals techniques such as the Grimme's method used here \cite{Anderson13, Klimes10, Carrasco11}. A direct comparison with experiment to confirm the accuracy of the van der Waals technique used is not possible due to the fact that the only available early experiment results, with adsorption energy estimates in the 1-2 eV range, \cite{Broden79, Benziger80, Wedler82} are not site-specific and not accurate enough to allow us to distinguish results at the $\sim$10 meV scale necessary here.

%%%%%%%%%%%%%%%%%%%%%%%%%%%%  TABLE ADSORPTION ENERGY 

\begin{table}[ht]
\caption{\label{Ads_eng} Adsorption energies of CO on different symmetric adsorption sites on the Fe-110 surface for the three coverages discussed in the text with and without van der Waals interaction. All energies are given in eV. } 
\vspace{2mm}
\setlength{\tabcolsep}{1mm}
\centering
\begin{tabular}{lccc}
\hline
\hline
Coverage (ML)	& 	On-top &	Hollow  &	Bridge	\\
\hline
			&  	\multicolumn {3}{c}{Without van der Waals interaction} \\
\hline

0.25			&      -1.93	&   -1.87	&    -1.66	\\
0.1111		&      -1.98	&   -1.92	&    -1.71	\\
0.0625	 	&      -2.07	&   -2.05	&    -1.88	\\

\hline
			& 	\multicolumn {3}{c}{With van der Waals interaction} \\
\hline
0.25			&      -2.11	&   -2.04	&   -1.87	\\
0.1111		&      -2.11    &   -2.05   	&   -1.88	\\
0.0625		&      -2.12	&   -2.09	&   -1.90	\\
				
\hline

\end{tabular}

\end{table}

%%%%%%%%%%%%%%%%%%%%%%%%%%%%%%%%%%%%%%%%%%%%%%%%%

\subsection{Dissociation}

Once the CO molecule is adsorbed on the surface, it decomposes into surface-adsorbed C and O atoms and the C atom diffuses into the Fe surface to complete the carburization process. The dissociation path is estimated using the nudged elastic band (NEB) method.  For the 0.25ML, van der Waals interactions are not used to allow a direct comparison with previous works. For 0.1111 ML and  0.0625 ML, although we have shown that van der Waals interactions have little influence, we have included them to capture the most precise details along the minimum-energy reaction pathway (MEP).
 We select the on-top configuration as the initial configuration for the NEB calculation because it is the most favorable  adsorption site as shown in the previous section. The final configuration consists of  the CO molecule completely decomposed into C and O adatoms on the surface. According to earlier investigations~\cite{Erley81,Jiang04}, upon decomposition, the C atom is adsorbed on the hollow site (Figure \ref{sites}) with 5 Fe neighbors (4 in the surface layer and 1 in the 2$^{nd}$ layer). We show that the C atom penetrates 0.2-0.3 \AA{} deep in the surface. The O atom, on the other hand, hovers 0.7-0.9 \AA{} above the nearest-neighbor hollow site with a coordination number 4.
 However, it is interesting to note that at a lower coverage of 0.0625 ML, the O atom relaxes further from the hollow site to a so-called quasi-threefold site \cite{Jiang04} with a coordination number of 3. This can be observed in the data for Fe-O distance given in Table \ref{newtab}. 
This distortion is caused by the relaxation of the local atomic structure under a reduced interaction between the oxygen atoms and its periodic images once the surface supercell size is increased from 2$\times$2 to 4$\times$4 and has a significant contribution in lowering the dissociation barrier. A detailed discussion is made in the next section.
 For the 2$\times$2 (0.25 ML coverage) surface in the dissociated state, the C atom is adsorbed 0.21 {\AA} below the surface while the O atom sits at 0.71 {\AA} above the surface. This final configuration remains stable for the three coverages. However, the C atom sinks slightly deeper, 0.23 \AA{} and 0.3 \AA{} below the surface, for the 3$\times$3 (0.1111 ML) and 4$\times$4 (0.0625 ML) surface systems, respectively.
 The energetics  of this final configuration (D) is strongly affected  by the CO coverage ratio and the size of the simulation box. For the 2$\times$2 surface supercell, the fully relaxed dissociated configuration is 0.46 eV higher than the initial on-top bound configuration, in agreement with earlier reports by Jiang \textit{et al} \cite{Jiang04}.
 Interestingly,  the situation is reversed for the larger 3$\times$3 and 4$\times$4 supercells. For these two coverages the total  energy of the dissociated CO configuration  is found be 0.09 eV and 0.72 eV \emph{lower}, respectively, than that of the initially adsorbed CO, corresponding to lowest energy point on the whole reaction path.

\begin{figure}
\centering
\includegraphics[width=8.5cm]{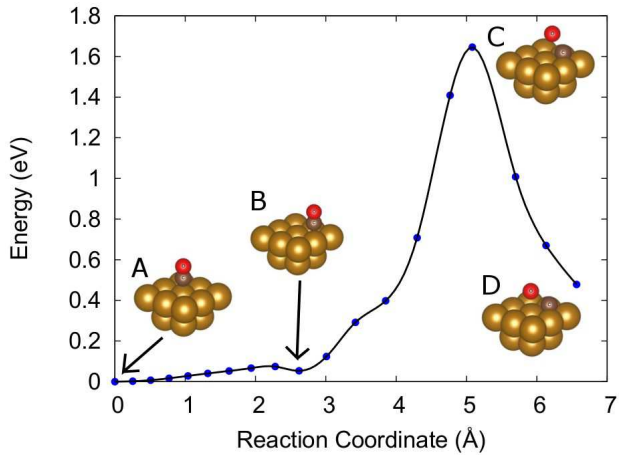}

\caption{Minimum energy path for CO-dissociation on clean Fe-110 surface with 2$\times$2 supercell (0.25 ML coverage), starting from on-top configurations. Important intermediate configurations are shown. A. Initial minimum: on-top adsorbed CO, B: Local minimum: Hollow site-adsorbed CO, C: Saddle point, CO bond severing, D: Decomposed into surface-adsorbed C and O atoms. }
\label{mep2x2}
\end{figure}

\begin{figure}
\centering
\includegraphics[width=9cm]{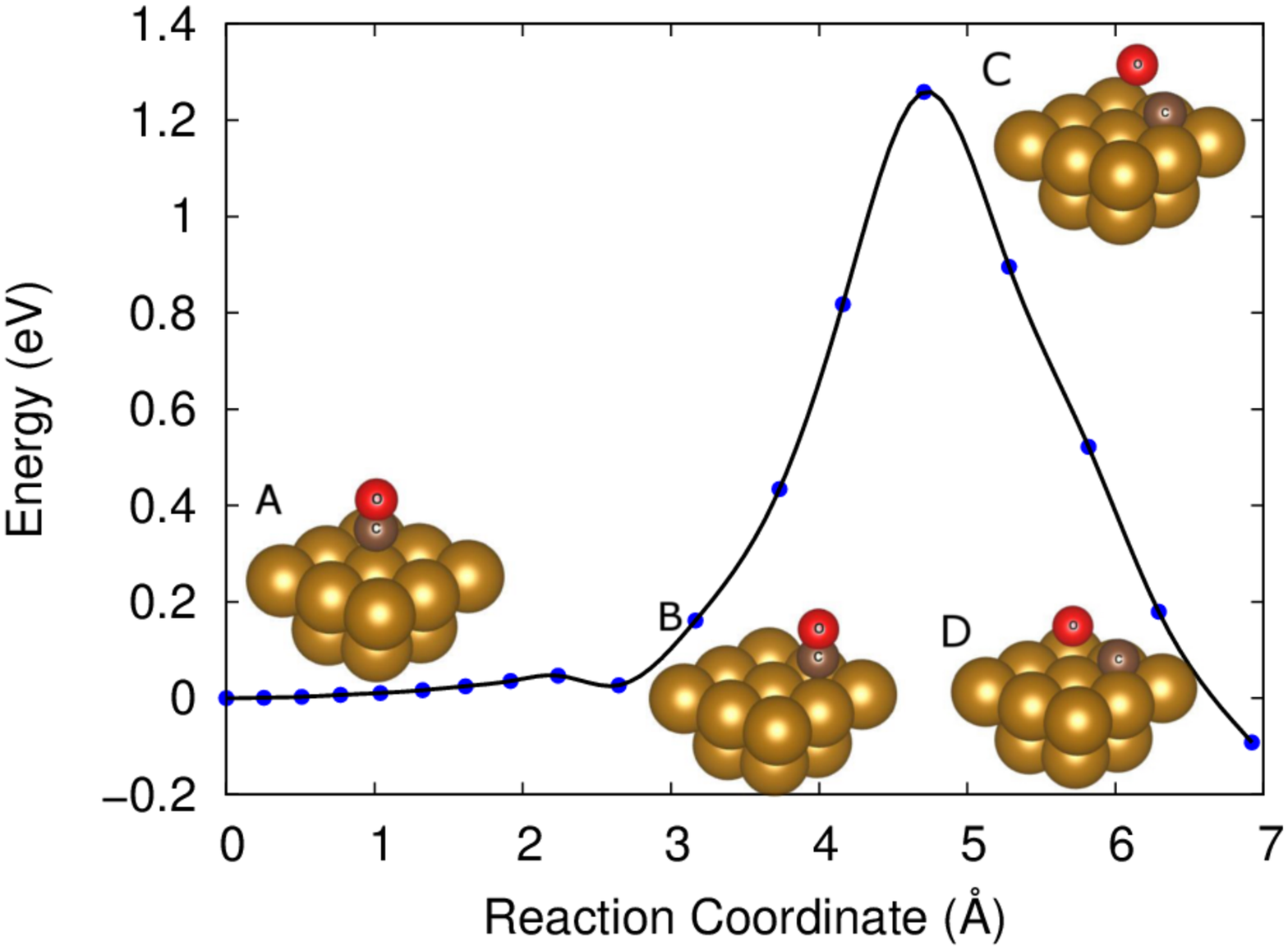}
\caption{Minimum energy path for CO dissociation with 3$\times$3 (0.1111 ML coverage).  Initial and final minima A, D and the intermediate points B, C are the same as in Fig. \ref{mep2x2}. The initial part A$\rightarrow$B is similar to that of a 0.25 ML system but the primary dissociation barrier (C) is reduced to 1.23 eV.} 
\label{mep3x3}
\end{figure}

\begin{figure}
\centering
\includegraphics[width=9cm]{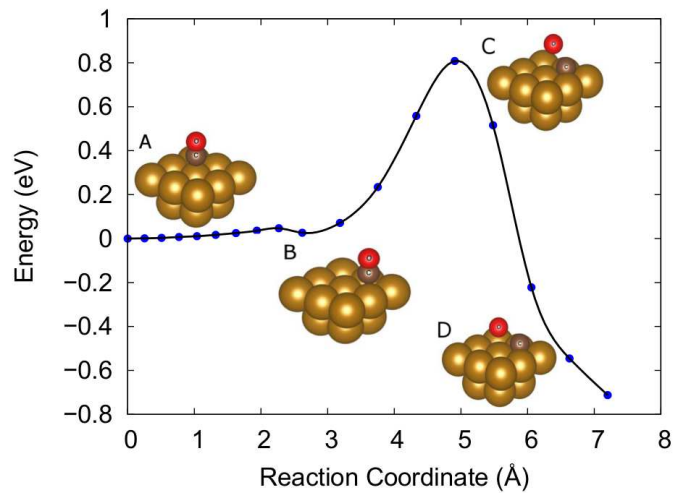}
\caption{Minimum energy path for CO dissociation with 4$\times$4 (0.0625 ML coverage).  Initial and final minima A, D and the intermediate points B, C are the same as in Fig. \ref{mep2x2}. The initial part A$\rightarrow$B is similar to that of a 0.25 ML system but the primary dissociation barrier is decreased to 0.78 eV. }
\label{mep4x4}
\end{figure}

Turning to the full MEP for the 2$\times$2  (0.25 ML) surface, the hollow site-adsorbed CO serves as an intermediate minimum-energy state (Configuration B in Fig. \ref{mep2x2}) in the reaction path from on-top site-adsorbed to fully dissociated state (A$\rightarrow$D in Fig. \ref{mep2x2}). In Table \ref{Ads_eng} we see that the hollow site-adsorbed CO has an energy slightly higher by 0.06 eV for 0.25 ML and 0.1111 ML and 0.03 eV for 0.0625 ML. Therefore we can split the reaction path into two segments. In the first, migration of the CO molecule as a whole takes place from the on-top to hollow adsorption sites, corresponding to path A$\rightarrow$B  in Figures \ref{mep2x2}, \ref{mep3x3} and \ref{mep4x4}. In the second,  we observe a complete decomposition of the molecule from the hollow site to the final dissociated configuration, as represented by B$\rightarrow$D. 
For the 1st segment, we find a similar minimum energy path  for all coverages with a small barrier of 0.07 (0.25 ML) - 0.05 (0,0625 ML) eV. For the second segment  B$\rightarrow$D, the primary barrier occurs at the saddle point C where the C-O bond is stretched to 2.1 {\AA} before  breaking to separate atoms. For the 2$\times$2 (0.25 ML) system we observe a barrier of 1.64 eV, which is comparable to that obtained by Jiang \textit{et al} \cite{Jiang04}. Fig. \ref{mep2x2} shows the complete minimum energy path for 0.25 ML, also in agreement with the literature \cite{Jiang04}.
 The first segment of the  MEP is similar for all surface supercells/coverages, leading to a metastable state at B,
the 2$^{nd}$ segment (B$\rightarrow$D) of the MEP becomes distinctively different. The primary saddle point is reached, at around 5
{\AA} from the initial minimum, for all coverages, leading to the breaking of the C-O
bond followed by a  sharp descent to the dissociated state. The energy associated with this primary barrier represents the dissociation barrier and are 1.23 eV and 0.78 eV for 3$\times$3 (0.1111 ML) and 4$\times$4 (0.0625 ML) systems, respectively (Figures \ref{mep3x3} and \ref{mep4x4}), 
showing a systematic reduction as the surface size increases. 
Important parameters related to the atomic structures of the initial, final, intermediate states and the saddle points along the MEP for all different supercells are shown in Table \ref{newtab}. A detailed analysis about the finite size effects and their roles in the systematic reduction of the dissociation barrier with increasing system size is presented in the Discussion section.

%%%%%%%%%%%%%%%%%%%%%%%%%%%%%%
\begin{widetext}

\begin{center}
\begin{table}[ht]
\caption{\label{newtab} Details of the local environments of the C and O atoms at different points along the dissociation path: relative energy (E$_S$) with respect to the initial on-top adsorption state, separation between periodic images d$_{CO-CO}$, bond-lengths d$_{C-O}$, d$_{Fe-C}$, d$_{Fe-O}$ and depth (vertical distance) of the C atom from the surface (d$_C$). The numbers in parentheses are from calculations \emph{with} van der Waals interaction. Note that the change in bond-lengths are small when van der Waals interaction is included and are not shown for atomic C+O states as the differences are insignificant.}
\vspace{2mm}
\setlength{\tabcolsep}{1mm}
\centering
\begin{tabular}{lccccc}
\hline
\hline

			& 2x2 (0.250 ML)	 &  2x3 (0.1666 ML) &   3x3 (0.1111 ML)	&  3x4 (0.0833 ML)	&  4x4 (0.0625 ML)	\\
\hline	
d$_{CO-CO}$ (\AA) & 4.9		& 4.9		& 7.4 		& 7.4		& 9.8		\\
\hline
&			\multicolumn {4}{c}{Initial state (A)}    \\
\hline
 d$_{C-O}$ (\AA)	 & 1.24 (1.22)	& 1.23 (1.22)	& 1.21 (1.20)	& 1.21 (1.20)	& 1.19 (1.19)	\\
d$_{Fe-C}$	(\AA)	&  1.10 (1.00)	& 1.10 (1.00)	& 0.98 (0.98)	& 0.96 (0.95)	& 0.91 (0.91)	\\

\hline	
&			\multicolumn {4}{c}{Intermediate minimum (B - hollow site adsorbed CO)}    \\
\hline
E$_S$ (eV)		 & 0.06 (0.07)	& 0.06 (0.07)	&  0.06 (0.06)	& 0.05 (0.05)	& 0.03 (0.02)	\\
 d$_{C-O}$ (\AA)	 & 1.25 (1.23)	& 1.24 (1.23)	& 1.23 (1.21)	& 1.21 (1.21)	& 1.19 (1.19)	\\
d$_{Fe-C}$	(\AA)	&  1.97 (1.96)	& 1.97 (1.96)	& 1.96 (1.96)	& 1.96 (1.96)	& 1.95 (1.95)	\\

\hline	
&			\multicolumn {4}{c}{Transition state (C)} 		\\
\hline
E$_S$ (eV)		&  1.64 (1.69)	& 1.51 (1.55)	& 1.19 (1.23)	& 1.16 (1.18)	& 0.78 (0.78)	\\	
d$_{C-O}$ (\AA)	&  2.04 (2.04)	& 2.05 (2.06)	& 2.07 (2.07)	& 2.09 (2.09)	& 2.15 (2.15)	\\
d$_{Fe-C}$	(\AA)	&  1.80		& 1.80		& 1.83		& 1.83		& 1.85		\\
d$_{Fe-O}$	(\AA)	&  2.04		& 2.01		& 1.98		& 1.97		& 1.91		\\
d$_C$		(\AA)  &  0.04		& 0.04		& 0.09		& 0.10		& 0.16		\\	

\hline
&			\multicolumn {4}{c}{Dissociated state (D)}    \\
\hline
E$_S$ (eV)		 & 0.46 (0.61)	& 0.39 (0.53)	&  -0.12 (-0.09)	& -0.19 (-0.11)	& -0.73 (-0.72)	\\
d$_{C-O}$ (\AA)	 & 2.74		& 2.81		& 2.93		& 2.96		& 3.14		\\
d$_{Fe-C}$	(\AA)	 & 1.77		& 1.77		& 1.78		& 1.78		& 1.79		\\
d$_{Fe-C}^{\prime}$ (\AA) & 1.87	& 1.88		& 1.92		& 1.92		& 1.96		\\
d$_{Fe-O}$	(\AA)	&  1.97 (1.95)	& 1.96 (1.95)	& 1.93 (1.92)	& 1.93 (1.92)	& 1.82 (1.82)	\\
d$_C$		(\AA)  &  0.21		& 0.21		& 0.23		& 0.24		& 0.30		\\	

\hline

\end{tabular}
\end{table}
\end{center}
\end{widetext}

 A lower dissociation barrier along with a
 slightly lower adsorption energy (See Table \ref{Ads_eng}) for low coverage systems imply
that the reaction between Fe-110 surface and CO is more favorable in a dilute coverage
of CO. Note that the reverse barrier for this reaction reaches 1.5 eV indicating that the
possibility of surface-embedded C to react with O to form CO is much less favorable.
Furthermore, previously published \ab results show a surface-to-subsurface diffusion
barrier of 1.2 eV for C at a coverage of 0.1111 ML \cite{Jiang05} which is less than the
reverse barrier of 1.35 eV for dissociation at the same coverage and much less than that at 0.0625 ML, 1.5 eV.
 Since structural relaxation is less constrained
in 0.0625 ML system than for a 0.1111 ML, it can be intuitively predicted that the
surface-to-subsurface diffusion barrier is even lower than 1.2 eV for 0.0625 ML system
and it is energetically favorable for the surface-embedded C atom to diffuse into Fe-110
subsurface than to react back with O to form CO.

 A similar study made on 111 surface \cite{Booyens14} shows that the adsorption energies are in the similar range as those for the 110 surface, and it presents a 1.5 eV dissociation barrier. However, a 0.25 ML coverage is used in this work and therefore is assumed to be not free from the restricted relaxation problem. The dissociated-CO state is reported to be 
$\sim$ -0.15 eV lower than the adsorbed-CO state, which is deeper than that for 110 surface, at the same 0.25 ML coverage (+0.46 eV) but is shallower than that at 0.0625 ML (-0.72 eV). The deeper CO-dissociated state energy is a result of the open structure of the 111 surface. However, looking at the significance of the finite-size effect presented in our work, a prediction can be made that the dissociated state may be even deeper for a dilute coverage / larger surface and may present a low dissociation barrier in the 111 surface as well.

\section{Discussion}
\label{sec:discussion}

\subsection{Role of finite-size effects on \emph{adsorption}}

\begin{figure}
\centering
 \includegraphics[width=9.5cm]{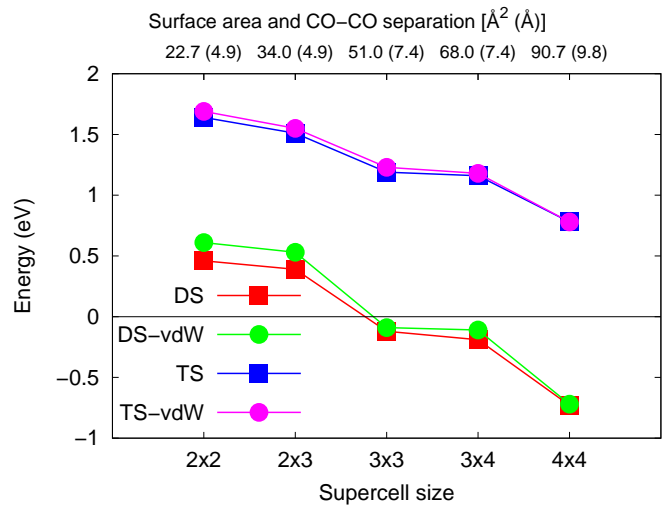}
\caption{Relative energies of the dissociated states (DS) and transition states (TS) with respect to the adsorbed state with (-vdW) and without van der Waals plotted as functions of coverage (supercell size). One CO atom in surface supercells 2$\times$2, 2$\times$3, 3$\times$3, 3$\times$4 and 4$\times$4 corresponds to coverages 0.25, 0.1666, 0.1111, 0.0833 and 0.0625 monolayers, respectively. Also shown are the surface areas for these supercells with minimum molecular distances for adsorbed CO molecules in parentheses. The minimum energy adsorption state, i.e. the on-top site adsorbed configuration, is considered for all systems.}
\label{newfig}
\end{figure}

  Adsorption at a high coverage is affected by the van der Waals interaction due to the strong dipole moments possessed by the CO molecule. Table \ref{Ads_eng} shows that the adsorption energy is heavily modified for the 2$\times2$ (0.25 ML coverage) system when van der Waals interactions are taken into account.
 However, for the 4$\times$4 110-surface supercell features a distance of 9.8 {\AA} between periodic images of CO molecules, the effect of van der Waals interactions is reduced by one order of magnitude to only 20-50 meVs.
 Note that we may require an even larger supercell to completely ignore the finite size effects in adsorption due to van der Waals interactions. 
As the CO molecule dissociates into surface-embedded C and O atoms, the van der Waals interactions are weakened due to the lack of dipole moment. They become therefore much less relevant in the dissociation process.
 This can be seen in Figure \ref{newfig} where the relative energies of the dissociated states with respect to the adsorbed state are plotted against the supercell size with and without van der Waals interaction.
 A clear pattern emerges: for the smaller supercells corresponding to higher coverages, the relative energy of the dissociated states are higher when van der Waals interaction is included. This behavior is not caused by the van der Walls interactions. It can be explained rather by  the modification in the energy of adsorption state with respect to which the relative energy is calculated with.
This difference decreases systematically as the system size --- and therefore the CO-CO periodic image separation--- increases. The energy changes of the dissociated states upon inclusion of van der Waals are minuscule due to the lack of dipole moment that was provided by the CO molecule for the adsorbed state. At the transition state, in particular, van der Waals interactions are barely noticeable. Other finite-size effects also  contribute to the adsorption energy: the lattice deformation of the surface around the adsorbed CO molecule. However, for a densely packed surface such as 110, this deformation is less than 10\% and is limited to the 1$^{st}$ nearest neighbor. 

\subsection{Finite-size effect on \emph{dissociation}}

 In order to understand the significance of the large finite size effect on the dissociation of CO, we study and analyze the factors that are related to the periodic boundary conditions and dictate the energy change in the transition state.
Figure \ref{newfig} depicts the transition state energies relative to the initial minimum.
 This energy represents the energy barrier for dissociation and we can see that it follows the same pattern as the relative energy of the final minimum corresponding to the dissociated state.
 We know that the transition state represents the C-O bond severing by catalytic reaction with Fe \cite{Jiang04, Elahifard12, Booyens14}.
 From an electronic structure perspective, there are two simultaneous mechanisms that enable the C-O bond severing: 1) binding of the C atoms with strong covalent bonds to the most stable configuration (in this case hollow site) on the Fe surface and 2) compensation of the high electron affinity of the O atom by charge transfer from the surface to the O atom \cite{Roncancio15}. 
Here, we focus on the effect of the periodic boundary conditions on these two mechanisms. Regarding the C-Fe bond formation,  Table \ref{newtab} shows that at the transition state, the C atom is already embedded in the Fe lattice (d$_C$ = 0.04 - 0.16 \AA), and it travels deeper in the dissociated state (d$_C$ = 0.21 - 0.3 \AA).
 Furthermore, for larger systems, the lattice  can relax further, providing a deeper penetration of the C atom to a more stable atomic position (Table \ref{newtab}) with higher Fe-C electronic overlaps, eventually decreasing the energy threshold for the C-O bond severing.
 Once the C  atom is embedded into the surface, it creates a large deformation in the Fe lattice, which interacts with its periodic images. This interaction, namely C-C interaction, is well known and there are a number of contributing factors to this interaction given by: 1) the short range intermolecular electrostatic interaction,  2) medium and long range elastic deformation, and 3) long range contributions due to a magnetic anisotropy \cite{Bhadeshia04}.
 The first factor is a component of the van der Waals interactions and we already discussed it to be a purely surface phenomenon. Electrostatic interactions between periodic images of  C  and O surface adatoms are short ranged in the dense Fe lattice and vanish within the first nearest neighbor. The magnetic contribution is commonly orders of magnitude weaker than the elastic contribution \cite{Seletskaia05} and can be ignored as well. Medium-range elastic interactions are the most relevant for the coverages we are studying here \cite{Bhadeshia04, Mou89}.

\begin{figure}
\centering
\includegraphics[width=6cm]{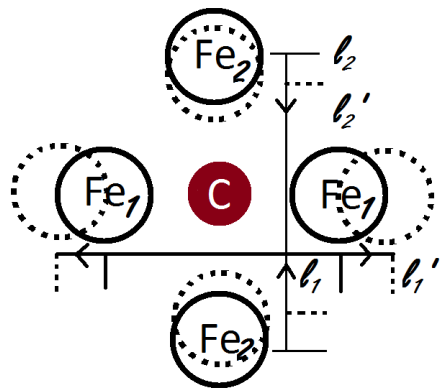} 
\caption{Deformation of a Fe lattice upon C insertion in a surface hollow site or an octahedral interstitial site. The solid circle shows two different symmetric Fe positions 1 and 2 and the dotted circles depicts their respective displaced positions upon C insertion. The distance between primary nearest neighbors Fe$_1$ atoms $l_1$ across the C site increases to $l_1^{\prime}$ while the secondary nearest-neighbor distance $l_2$ is decreased to $l_2^{\prime}$. Percentage change in these two distances are given by the deformation factors $\delta_1$ and $\delta_2$, where  $\delta_i = \frac{(l_i^{\prime} - l_i)}{l_i}\times 100, i = 1,2$}
\label{deformation}
\end{figure}

Detailed analysis of C-C interaction in $\alpha$-iron provided by Mou and Aaronson \cite{Mou89} shows that the elastic interaction is in a inverse-cubic relation with the distance between the two carbon interstitial atoms. These analytical studies were made for C interstitial in bulk iron but the nature of C-C interaction is expected to remain similar as we have shown in Table \ref{deform} that the average deformation caused by C insertion in the surface is comparable to that in a bulk. 
The C-C distance in the periodic supercells used here for 0.25 ML and 0.0625 ML coverages, \textit{i.e.} $2\times2$  and $4\times4$, are 4.9 and 9.8 \AA, respectively, which is a factor of two difference. Therefore, according to the inverse-cubic relation, the C-C interaction in the larger supercell/lower coverage is one order of magnitude smaller than that in the smaller supercell/higher coverage, leading to significant energy and structural differences between the two cell sizes. In order to compare the amount of deformation upon C-insertion in the two systems with bulk, we define the relative deformation factors  which are the percentage change in the Fe-Fe distance across the inserted C atom. Note that the surface hollow site, where the C atom is adsorbed, is equivalent to an octahedral interstitial site in bulk bcc structure. All octahedral interstitial sites as well as the surface hollow site have two different distances between the Fe neighbors surrounding it. This is represented by a schematic diagram in Fig. \ref{deformation}. If the distance between the Fe atoms are $l_1$ and $l_2$ across the octahedral cage, then upon C-insertion, the shorter distance $l_1$ expands to $l_1^{\prime}$  while the longer $l_2$ slightly shrinks to $l_2^{\prime}$. The deformation factors are defined by the percentage change of this distance.

\begin{table}[ht]
\caption{\label{deform} Changes in  the Fe-Fe distances upon C insertion  into the 110 surface and in the bulk of $\alpha$-Fe. A Fe-C surface system is chosen with one C atom in the most stable hollow site, and is equivalent to the dissociated state. Systems with increasing sizes are compared as reflected into the distance of C and its periodic images (d$_{C-C}$). $\delta_1 $  and $\delta_2$ represent the percentage change of the two symmetric  Fe-Fe distances (see Fig. \ref{deformation}) for the nearest neighbors of the C adatoms for surface or interstitial for bulk. d$_C$ represents the vertical depth of a C atom from the surface.}
\vspace{2mm}
\setlength{\tabcolsep}{1mm}
\centering
\begin{tabular}{lcccc}
\hline
\hline
Supercell &  d$_{C-C}$ (\AA)&	$\delta_1 (\%)$ & 	$\delta_2 (\%)$ &	d$_C$ (\AA)\\
\hline

2x2		&	4.9	&   +23.5	& -7.1		&	0.23 \\
2x3		&	4.9	&   +23.6	& -6.9		&	0.23 \\
3x3		&	7.4	&   +24.1	& -5.1		&	0.26 \\
3x4		&	7.4	&   +24.1	& -5.0		&	0.27 \\
4x4		&        9.8	&   +24.4	& -3.9 	&	0.33 \\

\hline
 		&	\multicolumn {3}{c}{Bulk}    \\
\hline
2x2x2			&     5.7	&   +22.1	&   -2.4 &	\\
3x3x3			&     8.5         &  +23.9	&   -2.2   & \\
4x4x4			&     11.4	&  +24.3\cite{Domain04} & -1.8 \cite{Domain04} &\\
				
\hline
\end{tabular}

\end{table}

Table \ref{deform} shows that there are large displacements in the primary nearest neighbors of the C atom, Fe$_1$ (see Fig. \ref{deformation}),  increasing the distance across the adatom/interstitial site 23-24\% of the distance.
 On the other hand, distance between the second nearest Fe- neighbors  (Fe$_2$) decreases. For the surface systems, the contractions in the second nearest neighbors (Fe$_2$ in Fig. \ref{deformation}) are larger.
 However, we note that deformations are not unlike that for the bulk and the amount of deformations in the $4\times4$ supercell are close to that of the large bulk systems.
 It is expected that the deformation should be larger for the surface adatoms than for bulk interstitials due to the presence of more degrees of freedom.
 A systematic convergence in $\delta_i$ can be observed, which is almost linear with the supercell' sizes with isotropic C-coverage ( $2\times2$, $3\times3$, and $4\times4$). 
  Note that the systems described in Table \ref{deform} are similar to the dissociated state and the converging pattern of the deformations is directly reflected in the relative energy of the dissociated states shown in Figure \ref{newfig}.
 In order to put the amount of deformation into perspective we compare with a larger $4\times4\times4$ bulk system with 128 atoms and a C interstitial (from literature \cite{Domain04})
 These deformations are large when applied to smaller periodic systems. Restricted from propagating, the uncompensated elastic energy results in a higher energy state and is precisely one of the major reasons of higher transition and dissociated states for the small systems.

As we identify the surface-embedded C-C interaction as one of the major contributor for reducing the dissociation barrier in dilute coverage Fe-CO systems, we must remember that a similar interaction occurs in any metal-hydrocarbon reaction where one or more adsorbate atom dissociate to get embedded in the host surface lattice. Therefore the finite-size effect we observe may have a significant impact in a wider field of study involving a similar reaction mechanism and it is highly recommended that a study using large surface systems with dilute coverage must be pursued to understand the fundamental electronic and atomic mechanisms for all such reactions. This bias is also present in the study of the influence of external adsorbates and defects on the reaction such that the effect of the external adsorbates are understood as isolated processes, free of finite-size effects. Here also, a higher-coverage study must accompany an estimate of the size effects such that the contribution of the reactants and finite-size effects can be separated.

The role of the oxygen atom at the transition state is also impacted by the finite-size effects. Table \ref{newtab} shows that for the larger systems the C- O distances in the transition and dissociated states are larger, while the Fe-O distance becomes smaller.
A better relaxed lattice allows the O atom to move closer to the surface, which enables a better charge transfer to compensate the high electron affinity of the O atom. This enables the C-O bond severing at a lower energy.
 As mentioned earlier, for the 4$\times$4 surface system, the O atom eventually moves from the hollow site (d$_{Fe-O}$ =1.96 \AA) to the less symmetric quasi-threefold site (d$_{Fe-O}$ =1.82 \AA) with 3 close-packed Fe neighbors. This site offers a lower energy and a better charge compensation for the O adatom.
Also the large electron affinity of the O atom restricts the C atom from forming strong covalent bonds with the Fe atoms. Better relaxed larger systems allows the O atom to displace further from the C atom, therefore rendering both the transition and dissociated states more stable.

It is inferred that at the low 0.0625 ML coverage associated with the 4$\times$4 surface supercell, the effect of periodic boundary condition or finite size effects  is considerably reduced.
 However, we must stress that the  $4\times4$ supercell is not completely free of finite size effects regarding both the van der Waals interaction in adsorption and C-C elastic interaction in dissociation of CO but is a sustainable compromise between accuracy and computational cost. High CO coverage systems can be modelled by adding more than one CO molecule on the 4$\times$4 surface supercell. With 4 CO molecules on the 4$\times$4 supercell, we have reproduced identical results with a 2$\times$2 supercell and one CO molecule representing the same coverage of 0.25 ML.This test confirms that the true origin of the size effects in adsorption and dissociation are CO-CO molecular or C-C elastic interactions, respectively. However, the large cell increases the number of degrees of freedom and allows lower symmetries that can also contribute to reduce elastic effects.
 This overall behavior is an important question and is particularly important if the effect of an external influence such as adsorbates or defects on the CO-Fe reaction are studied using a single supercell. The nature of finite-size effects described here are very general and may be applied to a number of other similar metal-hydrocarbon reaction systems but is yet unknown. We are currently working to address these questions and hope to come up with a more generalized picture in the near future.
 For a more accurate description of the hydrocarbon dissociation on complex metal surface systems including impurities and defects we are also working on developing semi-empirical and on-the-fly kinetic Monte-Carlo techniques alongside DFT to search for the minimum-energy dissociation paths.

\section{Conclusions}
Using PBE-GGA functional in DFT we calculate adsorption and dissociation parameters for CO on Fe-110 surface. We show that the widely used 2$\times$2 110 surface supercell system associated with 0.25 ML coverage is strongly affected by finite-size effects. In a $2\times2$ supercell, inclusion of van der Waals interactions changes the adsorption energies by about  0.2 eV, an effect which was overlooked in previous studies. Moving to lower coverages of 0.1111 ML and 0.0625 ML using $3\times3$ and $4\times4$  surface supercells, respectively, allowed us to study adsorption and dissociation as isolated processes with significantly decreased size effects. For $4\times4$ supercells, the effect of van der Waals interactions is one order of  magnitude smaller than that in 2$\times$2. 
However, van der Waals interactions do not affect the dissociation: at the transition state, the CO molecule is decomposed and the source of van der Waals interaction, the large dipole moment of CO, vanishes.
 On the other hand, free relaxation of the C-adsorbed lattice becomes the crucial element in dissociation as at the transition state the C atom gets embedded in the Fe surface.
 A large surface supercell allows an unobstructed relaxation of the surface Fe atoms with the C atom inserted, which reduces the uncompensated elastic energies originated from the periodic boundary condition. Furthermore, the relaxed lattice allows the O adatom to a reside in a more stable chemical environment. As a result, a low-energy transition state appears, associated with a much lower dissociation barrier,  0.78 eV for 0.0625 ML compared to 1.64 eV for the 0.25 ML coverage. This indicates that in larger simulation cells, corresponding to lower coverage, the CO-Fe-110 reaction is significantly more favourable than that had been estimated before. However, a similar finite-size effect on dissociation may occur in any metal-hydrocarbon reaction where one or more adsorbate atoms are inserted in the host lattice and therefore it is of paramount importance to assess the size effects using large surface systems in order to understand the detailed mechanism of the reaction as an isolated phenomenon, before proceeding to understand the influence of defect and impurities on these processes.

\begin{acknowledgments}
The advanced computing facility of Texas A\&M University at Qatar is used for all calculations. This work is supported by the Qatar National Research Fund (QNRF) through the National Priorities Research Program (NPRP 6-863-2-355).
\end{acknowledgments}

%\appendix

%\section{A little f on appendixes}

%\nocite{*}
\bibliography{MDCBiB}% Produces the bibliography via BibTeX.

\end{document}